\documentclass[conference,onecolumn]{IEEEtran}
\IEEEoverridecommandlockouts
\usepackage{cite}
\usepackage{amsmath,amssymb,amsfonts}
\usepackage{algorithmic}
\usepackage{blindtext}
\usepackage{graphicx}
\usepackage{textcomp}
\usepackage{xcolor}
\usepackage{wrapfig}
\def\BibTeX{{\rm B\kern-.05em{\sc i\kern-.025em b}\kern-.08em
    T\kern-.1667em\lower.7ex\hbox{E}\kern-.125emX}}
\begin{document}

\title{The Benefits of Using the S-Band in Optical Fiber Communications and How to Get There\\
\thanks{The work was supported by UK EPSRC programme grant TRANSNET (EP/R035342/1). The authors thank S. Makovejs from Corning for the fiber data.}
}

\author{
  \IEEEauthorblockN{
    Daniel Semrau,
    Eric Sillekens,
    Robert I. Killey,
    and Polina Bayvel
  }
  \vspace*{1ex}
  \IEEEauthorblockA{
    \IEEEauthorrefmark{1}Dept. Electronic \& Electrical Engineering, University College London, WC1E 7JE London, U.K. 
  }
}

\IEEEspecialpapernotice{(Invited Paper)}

\maketitle\vspace*{-1ex}

\begin{abstract}
The throughput gains of extending the optical transmission bandwidth to the S+C+L-band are quantified using a Gaussian Noise model that accounts for inter-channel stimulated Raman scattering (ISRS). The impact of potential ISRS mitigation strategies, such as dynamic gain equalization and power optimization, are investigated. 
\end{abstract}
\vspace*{.3ex}


\section{Introduction}
Ultra-wideband optical transmission systems allow increased fiber link capacities, and are being used by network operators aiming to fully exploit their existing single-mode fiber (SMF) infrastructure. Currently, most optical communication systems are operating over the C-band (5~THz wide), with some high speed links being operated over the entire C+L band (10~THz wide). Experimental demonstrations of C+L band systems transmitting over transoceanic distances date back to 2013 \cite{Salsi2013transmission}. Ever since, the capacity of transoceanic C+L systems gradually increased as shown in Fig. 1 (after \cite[Fig. 1]{Ionescu2019Transmission}), mainly driven by advanced coded modulation and digital signal processing. Modern C+L band systems achieve an approximate twofold increase compared to their C-band only counterparts. While C+L band systems are a commercial reality and the demand for fiber capacity is still growing, expanding transmission bandwidths to the entire S+C+L-band (20~THz wide) is increasingly explored. Enabled by ultra-wideband amplification technologies, such as thulium-doped fiber amplifiers (TDFA) and semiconductor optical amplifiers (SOA), first S+C+L band transmission systems were recently demonstrated over a continuous bandwidth of 12.5~THz \cite{Renaudier_2017_f1c} and discontineous bandwidths of 16.4~THz \cite{Hamaoka_2018_usf} and 15.1~THz \cite{Puttnam_2020_059}. However, to predict the capacity benefits of the S-band in mature and commercially available S+C+L band systems in the future, analytical models can be utilized. Recently, the formalism of the Gaussian Noise (GN) model was extended to include inter-channel stimulated Raman scattering (ISRS), enabling accurate predictions of ultra-wideband transmission performance \cite{Roberts_17_cpo, Cantono_2018_oti, Semrau_2018_tgn}. Furthermore, a modulation format dependent closed-form formula was introduced enabling large scale system optimization and capacity estimations for arbitrary modulation formats~\cite{Semrau_2019_aca, Semrau_2019_amf,Semrau_2019_mfd}. 
\begin{wrapfigure}[16]{r}{0.515\textwidth}
   \centering
    \includegraphics[]{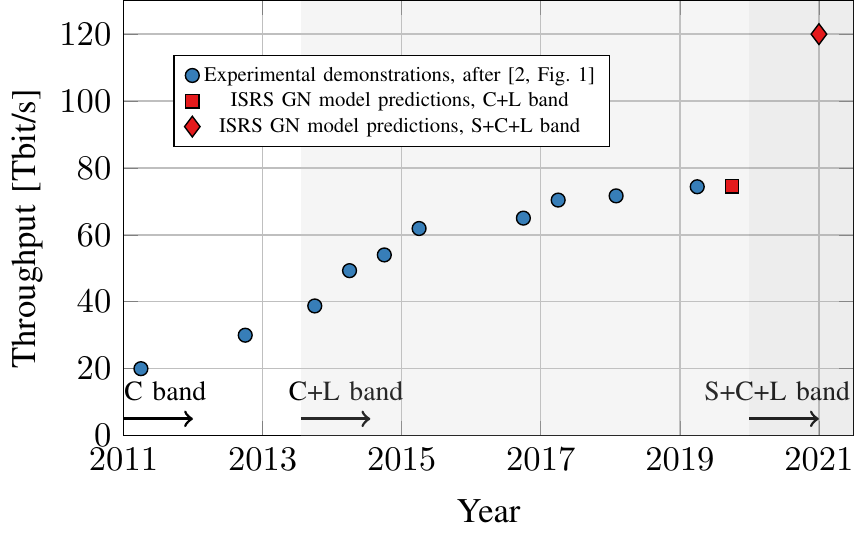}
\vspace{-0.1cm}
\caption{Records of transoceanic transmission experiments over $>5900$~km. Also shown, modelling predictions as a result of this work using the ISRS GN model.}
\label{fig:capacity}
\end{wrapfigure} 
\
In this paper, the throughput gains of extending the transmission window to the entire S+C+L band are analytically estimated using the modulation format dependent ISRS GN model. Two scenarios are studied; 1) using ideal dynamic gain equalization at each amplification stage, and 2) considering non-ideal gain equalization. The latter leads to ISRS accumulation across multiple fiber spans. Two distinct, geometrically shaped modulation formats and a finite set of available code rates across the optical spectrum are considered.  
\section{Analytical performance prediction}
%
To estimate the throughput of the system under test, we use a modulation format dependent, closed-form approximation of the ISRS GN model \cite{Semrau_2019_aca, Semrau_2019_amf}. The formula was derived assuming a triangular approximation of the Raman gain spectrum, valid up to 15~THz optical bandwidth. However in \cite{Semrau_2019_mfd}, we proposed a semi-analytical approach that enables the formula to be applied to bandwidths beyond 15~THz. The approach consists of matching the actual signal power profile to its first-order approximation (whose parameter are then used in the formula). The methodology yields very fast, but yet accurate, performance estimations and enables optimization problems with large sample space, such as launch power optimization. 
We consider 364$\times$50~GBd channels, centered at 1540~nm and transmitted over a transoceanic distance of 7000~km (100$\times$70~km). A transoceanic distance was chosen to ease the comparison to experimental demonstrations. Spectral gaps of 10~nm and 5~nm were considered between the respective transmission bands. A Corning\textsuperscript{\textregistered} SMF-28\textsuperscript{\textregistered} ULL fiber was assumed with dispersion and nonlinearity parameter as $D=18\frac{\text{ps}}{\text{nm}\cdot \text{km}}$, $S=0.067\frac{\text{ps}}{\text{nm}^2\cdot\text{km}}$ and $\gamma=1.2\frac{\text{1}}{\text{W}\cdot\text{km}}$ (see \cite{Semrau_2019_mfd} for more details). The noise figures of the EDFA/TDFA were 7~dB, 4~dB, and 6~dB in the S, C and L-band, respectively. The spectral launch power distribution was optimized using a particle swarm algorithm with subsequent gradient descent. Two distinct, geometrically shaped modulation formats were considered. In the S-band a 16-QAM format was transmitted (optimized for 7~dB SNR with excess kurtosis $\Phi=-0.49$), whereas in the C+L band a 64-QAM format was transmitted (optimized for 11~dB SNR with excess kurtosis $\Phi=-0.32$). Once the signal-to-noise ratio (SNR) was calculated, a finite set of code rates was selected. The code rates were optimized to maximize the total throughput and were upper bounded by the normalized generalized mutual information (GMI) of the respective channels.  
Two scenarios were considered: 1) ideal gain equalization, where the transmitted launch power was recovered at each span, and 2) non-ideal gain equalization, where after each span only 50\% of the ISRS power transfer is compensated and after every fifth span the transmitted launch power is ideally recovered. In the latter case, the ISRS power transfer is accumulating across multiple spans. The optimized SNR and launch power distribution are shown in Fig. 2a) and b), respectively. The inset in Fig. 2a) shows the considered modulation formats. The GMI and the total throughput depending on the number of code rates for both cases are shown in Fig. 2c). For comparison, the analysis was repeated for a C+L band system with the achieved throughput shown in Fig. 1. 
\begin{figure*}[t!]
   \centering
    \includegraphics[]{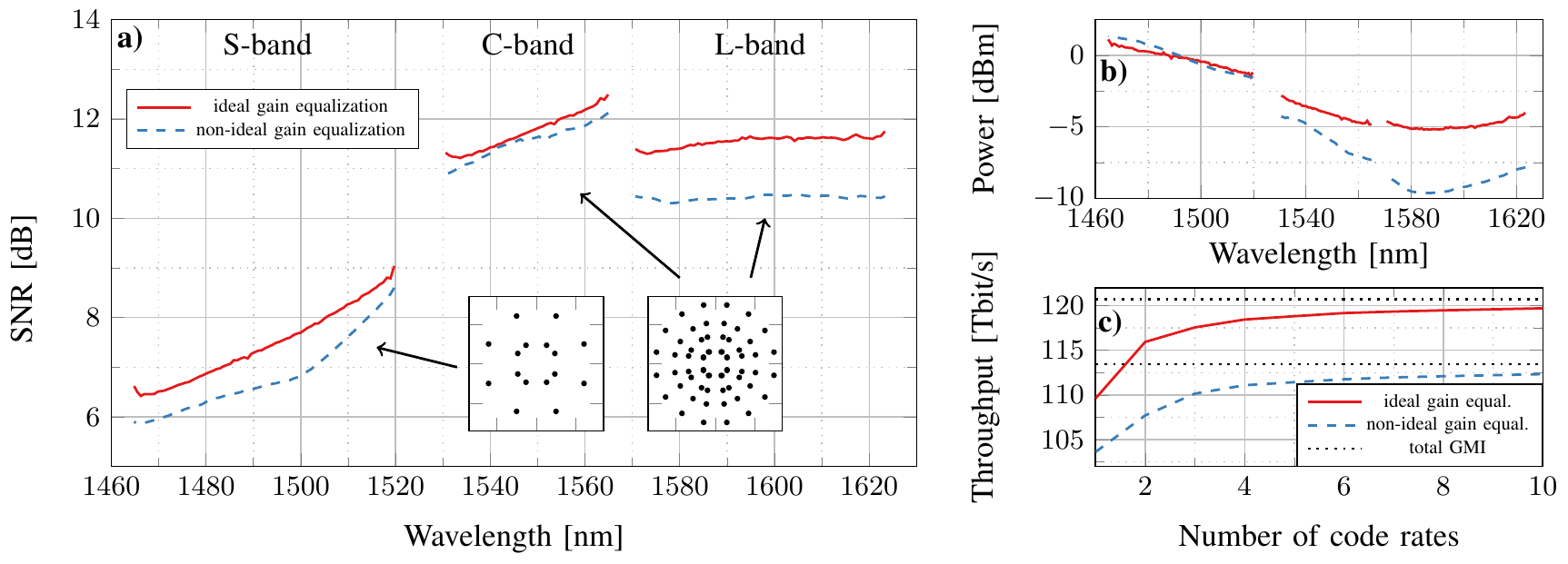}
\vspace{-1ex}
\caption{\small The a) SNR and b) launch power after $\text{100}\times\text{70~km}$ transmission for ideal and non-ideal gain equalization. The throughput and GMI vs. the number of code rates are shown in c). The insets in a) show the used modulation formats in the respective transmission bands.}
\vspace{-3ex}
\label{fig:NLI}
\end{figure*}
The results in this paper show that current throughputs can be increased from 74.59~Tbit/s to around 119.5~Tbit/s (achieving 119.2~Tbit/s using 6 code rates) when using the S-band for fiber transmission. This can be achieved by launch power optimization, using two distinct modulation formats for the S and C+L band and only a few code rates. It was also shown that non-ideal gain equalization reduces the throughput to around 112.3~Tbit/s, emphazing the importance of ISRS mitigation. Although the analysis in this paper was carried out for transoceanic distances, the conclusions carry over to terrestial distances. 

\section{Conclusions}
The benefits of using the S-band in optical communication systems were studied, indicating throughput gains of around 60\% compared with C+L band transmission. The capacity gain can be unlocked using different modulation formats in the respective transmission bands, spectrally optimized launch powers and dynamic gain equalization. 

\bibliographystyle{IEEEtran}
\bibliography{IEEEabrv,ref}

\begin{thebibliography}{10}
\providecommand{\url}[1]{#1}
\csname url@samestyle\endcsname
\providecommand{\newblock}{\relax}
\providecommand{\bibinfo}[2]{#2}
\providecommand{\BIBentrySTDinterwordspacing}{\spaceskip=0pt\relax}
\providecommand{\BIBentryALTinterwordstretchfactor}{4}
\providecommand{\BIBentryALTinterwordspacing}{\spaceskip=\fontdimen2\font plus
\BIBentryALTinterwordstretchfactor\fontdimen3\font minus
  \fontdimen4\font\relax}
\providecommand{\BIBforeignlanguage}[2]{{%
\expandafter\ifx\csname l@#1\endcsname\relax
\typeout{** WARNING: IEEEtran.bst: No hyphenation pattern has been}%
\typeout{** loaded for the language `#1'. Using the pattern for}%
\typeout{** the default language instead.}%
\else
\language=\csname l@#1\endcsname
\fi
#2}}
\providecommand{\BIBdecl}{\relax}
\BIBdecl

\bibitem{Salsi2013transmission}
M.~Salsi \emph{et~al.}, ``38.75 {Tb/s} transmission experiment over
  transoceanic distance,'' in \emph{Proc. ECOC}, 2013.

\bibitem{Ionescu2019Transmission}
M.~Ionescu \emph{et~al.}, ``74.38 {Tb/s} transmission over 6300 km single mode
  fiber with hybrid {EDFA}/{Raman} amplifiers,'' \emph{JLT}, \textbf{36}(1),
  2019.

\bibitem{Renaudier_2017_f1c}
J.~Renaudier \emph{et~al.}, ``First 100-nm continuous-band {WDM} transmission
  system with 115{T}b/s transport over 100km using novel ultra-wideband
  semiconductor optical amplifiers,'' in \emph{Proc. ECOC}, 2017.

\bibitem{Hamaoka_2018_usf}
F.~{Hamaoka} \emph{et~al.}, ``150.3-{T}b/s ultra-wideband ({S}, {C}, and {L}
  bands) single-mode fibre transmission over 40-km using $>$519{G}b/s/$\lambda$
  {PDM-128QAM} signals,'' in \emph{Proc. ECOC}, 2018.

\bibitem{Puttnam_2020_059}
B.~J. {Puttnam} \emph{et~al.}, ``0.596 {P}b/s {S}, {C}, {L}-band transmission
  in a 125$\mu$m diameter 4-core fiber using a single wideband comb source,''
  in \emph{Proc. OFC}, 2020.

\bibitem{Roberts_17_cpo}
I.~Roberts \emph{et~al.}, ``Channel power optimization of {WDM} systems
  following {Gaussian} {N}oise nonlinearity model in presence of stimulated
  {Raman} scattering,'' \emph{JLT}, \textbf{35}(23), 2017.

\bibitem{Cantono_2018_oti}
M.~Cantono \emph{et~al.}, ``On the interplay of nonlinear interference
  generation with stimulated {R}aman scattering for {QoT} estimation,''
  \emph{JLT}, \textbf{36}(15), 2018.

\bibitem{Semrau_2018_tgn}
D.~Semrau \emph{et~al.}, ``The {G}aussian {N}oise model in the presence of
  inter-channel stimulated {R}aman scattering,'' \emph{JLT}, \textbf{36}(14),
  2018.

\bibitem{Semrau_2019_aca}
D.~{Semrau} \emph{et~al.}, ``A closed-form approximation of the {G}aussian
  {N}oise model in the presence of inter-channel stimulated {R}aman
  scattering,'' \emph{JLT}, \textbf{37}(9), 2019.

\bibitem{Semrau_2019_amf}
{D. Semrau} \emph{et~al.}, ``A modulation format correction formula for the
  {G}aussian noise model in the presence of inter-channel stimulated {R}aman
  scattering,'' \emph{JLT}, \textbf{37}(19), 2019.

\bibitem{Semrau_2019_mfd}
D.~Semrau \emph{et~al.}, ``Modulation format dependent, closed-form formula for
  estimating nonlinear interference in {S+C+L} band systems,'' in \emph{Proc.
  ECOC}, 2019.

\end{thebibliography}

\end{document}